\documentclass[aps,prd,reprint,superscriptaddress,nofootinbib]{revtex4-2} %

\newcommand\aap{A\&A}                
\newcommand\aj{AJ}                   
\newcommand\aplett{Astrophys.~Lett.} 
\newcommand\apjl{ApJ}                
\newcommand\apjs{ApJS}               
\newcommand\icarus{Icarus}           
\newcommand\mnras{MNRAS}             
\newcommand\pasa{Publ. Astron. Soc. Australia}  
\newcommand\planss{Planet. Space~Sci.} 

\input

\newcommand{\REV}[1]{{{#1}}}

\usepackage{xcolor}
\usepackage{graphicx}	
\usepackage{amsmath}	
\usepackage{amssymb}	
\usepackage{hyperref}

\newcommand{\yr}{{\,\rm yr}}
\newcommand{\au}{{\,\rm au}}

\newcommand{\msun}{{\,\rm M_\odot}}
\newcommand{\rsun}{{\,\rm R_\odot}}

\begin{document}


\title{Revisiting Common Envelope Evolution -- A New Semi-Analytic Model for N-body and Population Synthesis Codes}


\author{Alessandro Alberto Trani}
\affiliation{Research Center for the Early Universe, Graduate School of Science,
The University of Tokyo, 7-3-1 Hongo, Bunkyo-ku, Tokyo 113-0033, Japan}
\affiliation{Okinawa Institute of Science and Technology, 1919-1 Tancha, Onna-son, Okinawa 904-0495, Japan}
\affiliation{Department of Earth Science and Astronomy, College of Arts and Sciences, The University of Tokyo, 3-8-1 Komaba, Meguro-ku, Tokyo 153-8902, Japan}

\author{Steven Rieder}
\affiliation{Geneva Observatory, University of Geneva, Chemin Pegasi 51, 1290 Sauverny, Switzerland}
\affiliation{School of Physics and Astronomy, University of Exeter, Stocker Road, Exeter, EX4 4QL, UK}

\author{Ataru Tanikawa}
\affiliation{Department of Earth Science and Astronomy, College of Arts and Sciences, The University of Tokyo, 3-8-1 Komaba, Meguro-ku, Tokyo 153-8902, Japan}

\author{Giuliano Iorio}
\affiliation{Dipartimento di Fisica e Astronomia “Galileo Galilei”, Università di Padova, vicolo dell’Osservatorio 3, IT-35122, Padova, Italy }
\affiliation{INAF - Osservatorio Astronomico di Padova, vicolo dell’Osservatorio 5, IT-35122 Padova, Italy}
\affiliation{INFN-Padova, Via Marzolo 8, I–35131 Padova, Italy}

\author{Riccardo Martini}
\affiliation{INFN - Sezione di Pisa, Edificio C – Polo Fibonacci Largo B. Pontecorvo, 3 – 56127 Pisa, Italy}
\affiliation{Okinawa Institute of Science and Technology, 1919-1 Tancha, Onna-son, Okinawa 904-0495, Japan}

\author{Georgii Karelin}
\affiliation{Okinawa Institute of Science and Technology, 1919-1 Tancha, Onna-son, Okinawa 904-0495, Japan}

\author{Hila Glanz}
\affiliation{Technion - Israel Institute of Technology, Haifa, 3200002, Israel}

\author{Simon Portegies Zwart}
\affiliation{Leiden Observatory, Leiden University, PO Box 9513, 2300 RA, Leiden, The Netherlands}
%


\date{\today}

\begin{abstract}



We present a novel way of modeling common envelope evolution in binary and few-body systems.
We consider the common envelope inspiral as driven by a drag force with a power-law dependence in relative distance and velocity.
The orbital motion is resolved either by direct $N$-body integration or by solving the set of differential equations for the orbital elements as derived using perturbation theory.
Our formalism can model the eccentricity during the common envelope inspiral, and it gives results consistent with smoothed particles hydrodynamical simulations.
We apply our formalism to common envelope events from binary population synthesis models and find that the final eccentricity distribution resembles the observed distribution of post-common-envelope binaries.
Our model can be used for time-resolved common-envelope evolution in population synthesis calculations or as part of binary interactions in direct $N$-body simulations of star clusters.
\end{abstract}


\maketitle

\section{Introduction}

Common envelope (CE) evolution is the process during which one component of a binary star gets engulfed in the envelope of its companion. During this phase, the gaseous envelope becomes gravitationally focused and exerts a drag force onto the stars, which begin to inspiral towards each other.
CE evolution ends when either the two stars merge or the envelope is (partially) ejected.
If the envelope is ejected and the two stars survive, the post-CE binary separation is much shorter than the initial one.
Originally proposed to explain the existence of short-period dwarf binaries \citep{paczynski1976,han2003a,han2003b}, CE evolution is now the key process of many other astrophysical phenomena, from gravitational wave sources \cite[e.g.][see \citealp{mapelli2021review} and references therein]{belczynski2007,dominik2015,belczynski2020,tanikawa2022}, to X-ray binaries \citep{kalogera1996,podsiadlowski2002}, and type Ia supernovae \citep{iben1984,webbink1984,han2004}.

CE evolution is arguably one of the least understood phases of interacting binary stars.
From the observational point of view, direct detection of CE is exceptionally elusive, first because of its short duration \REV{({a few years for the rapid inspiral phase, and possibly up to $10^5 \yr$ for the complete envelope ejection \citep{michaely2019b,igoshev2019,glanz2018}})}, and second because the binary is hidden by the CE, making it appear as a giant star.
Nevertheless, luminous red novae, a new class of transients, have been claimed as a promising candidate of CE events \citep{tylenda2011,ivanova2013b,macleod2017}.

Theoretical models of CE also have their share of limitations.
Hydrodynamical simulations are computationally expensive, and they either are not able to model the entire CE evolution or miss some physical ingredients \REV{(e.g., recombination of the envelope's gas, radiative and convective transport) \citep{sandquist1998,ricker2012,passy2012,ivanova2016,ohlmann2017,iaconi2018,reichardt2019,reichardt2020,vignagomez2020,glanz2021a,lau2022a,lau2022b}.} 
\REV{Similar considerations apply to 1D models of CE evolution, which can follow the slow, self-regulating evolution of the CE inspiral better than 3D hydrodynamics but miss other key aspects \citep{taam1978,taam1979,meyer1979,podsiadlowski2001,ivanova2002,ivanova2016,clayton2017,fragos2019}.} 
Conversely, all demographic studies of compact object formation adopt much simpler analytic formalisms \cite{vanDenHeuvel1976,paczynski1976,webbink1984,livio1988,dekool1990,nelemans2000,nelemans2005}.
\REV{One of the most widely adopted formalisms} is the $\alpha \lambda$ model, which is based on a simple energy balance equation that takes into account the binding energy of the envelope (parametrized by $\lambda$) and the efficiency of CE inspiral (parametrized by $\alpha$, see  \autoref{sec:alphalambdaconsist} for more details).
Because of its straightforward implementation and computational ease, most binary population synthesis (BPS) codes adopt the $\alpha \lambda$ model \citep[e.g.][]{portegieszwart1996,hurley2000,izzard2006,belczynski2008,eldridge2016,giacobbo2018,kruckow2018,spera2019,cosmic2020,compas2022,tanikawa2022}.

However, simplicity comes at a cost.
Being based on an energy balance equation, the $\alpha \lambda$ model completely neglects angular momentum. 
Consequently, it is not possible to meaningfully predict the post-CE eccentricity in BPS codes, which is always set to zero\footnote[1]{BPS codes like \textsc{bse} use the CE energy loss first to circularize the orbit. In principle, they may allow a final eccentric orbit if the energy loss is less than that required to circularize the binary. In practice, this is never the case when significant shrinking of the orbit occurs.}.
While most observations suggest that, in fact, CE events circularize the binaries, some observations of post-CE systems show residual, non-negligible eccentricities \citep{delfosse1999,edelmann2005,lynch2012,kawka2015,kruckow2021}.

Another drawback is that CE evolution in BPS codes is instantaneous, i.e., it consists of a jump in orbital separations from the pre-CE semimajor axis to the post-CE one.
In BPS codes, this makes it impossible to handle other processes that may occur during CE evolution, like supernovae explosions.
This issue is exacerbated when combining BPS codes with direct $N$-body codes \citep{aarsethnb7,petarcode} or secular evolution codes for multiple stellar systems \citep{toonen2016,hamers2021}, which poorly handle discontinuities in the evolution.

In this paper, we propose an alternative approach for CE evolution that can overcome these limitations and can be applied to $N$-body and BPS codes alike. 

The main assumptions of our model are presented in \autoref{sec:dragforceform}. 
Using perturbation theory, we derive the rate of change in orbital semimajor axis $a$, eccentricity $e$, and argument of pericenter $\omega$ due to the drag (\autoref{sec:forcepowlaw}). 
In \autoref{sec:halt} we discuss how the halting of the CE inspiral can be incorporated in our model, either by assuming self-similar expansion of the envelope or by estimating the energy losses with the $\alpha \lambda$ model. 
We compare our model with hydrodynamical simulations to find the most suitable form for the drag force in \autoref{sec:hydrocomp}. 
Finally, we apply our model to CE evolution triggered by the excitation of eccentricity in triple systems (\autoref{sec:CEkozai}), and to CE in isolated binary evolution (\autoref{sec:bspcomparison}). 
Comparing our model with the CE evolution in BPS codes, we find that our model predicts non-zero orbital eccentricities, similar to the observations. This and other results are summarized in \autoref{sec:concl}.

\section{Drag force formalism}\label{sec:dragforceform}

We consider that the two bodies undergoing CE evolution are experiencing a drag due to the surrounding gas.
We assume that the drag force is always opposite to the bodies' velocity vectors, and express it in the following general form:

\begin{equation}\label{eq:dragform}
	\mathbf{F} = - C \, v^l \, P(r) \,\hat{v}\,,
\end{equation}
where $C$ is a dimensional constant, $l$ is a real number that sets the drag force dependence on the relative velocity. $P(r)$ is a function representing the drag force dependence on radius. In this paper, we examine a power-law form for the function $P(r)$ (\autoref{sec:forcepowlaw}).

To better motivate the choice of the drag force, we compare \autoref{eq:dragform} with the expression for the fluidodynamical drag force:
\begin{equation}\label{eq:dragphys}
	F_\mathrm{drag} = -\frac{1}{2} \rho \, v^2 C_\mathrm{D} \, A\,.
\end{equation}

In the expression above, $\rho$ represents the local density of the fluid, $A$ is the cross-sectional area of the body immersed in the fluid, and $C_\mathrm{D}$ is a dimensionless coefficient that depends on the Reynolds and Mach numbers.
At high Mach numbers, $C_\mathrm{D} = 2$, while at low Mach and Reynolds numbers $C_\mathrm{D} \propto 1/v$, which makes the force scale as $F_\mathrm{drag} \propto v$.

\autoref{eq:dragphys} refers to the fluidodynamical drag force of a body immersed in a viscous fluid. \REV{On the other hand, the drag force during CE evolution is thought to be caused by the dynamical friction from the gravitationally focused gas, as supported by hydrodynamical simulations.
Numerical experiments have further shown that the gravitational drag is well expressed by \autoref{eq:dragphys}, albeit with a different dimensionless coefficient \cite{shima1985,macleod2017b,reichardt2019}.}

\autoref{eq:dragform} reduces to \autoref{eq:dragphys} if $l=2$ and $P(r) := A \rho$.
Therefore, the function $P(r)$ expresses the dependency of the gas density $\rho(r)$ and cross-sectional area $A(r)$ as a function of the distance between the two bodies.

If acceleration from the drag force is small compared to the mutual gravity, and once given an analytic expression for $P(r)$, we can apply the classical tools of perturbation theory and derive the evolution of the binary's orbital elements.

In the following sections, we choose an analytically convenient form for $P(r)$ and derive the corresponding differential equations for the orbital elements.
As a first approximation, we consider the two bodies to be point masses of mass $m_1$ and $m_2$, located at the center of mass of each star, corresponding to the stellar cores.
Mass loss and transfer can be added as extra terms in a second step \citep[e.g.][]{dosopoulou2016}.
For completeness, in the Appendix~\ref{sec:massloss}, we present the additional terms in $\dot{a}$, $\dot{e}$, $\dot{\omega}$ that describe the change in orbital elements under the assumption of isotropic mass loss.
Knowing that the force in \autoref{eq:dragform} has no component outside the plane of the binary, we can decompose the acceleration into its radial and tangential components:
\begin{equation}\label{eq:accelcomp}
	\mathbf{f} = f_r \hat{r} + f_\nu \hat{\nu}\,.
\end{equation}
From these we derive the time derivative of the binary semimajor axis $a$, eccentricity $e$, argument of pericenter $\omega$ and true anomaly $\nu$:
\begin{align}\label{eq:perturbeq}
	& \dot{a} = \frac{2 a^2}{\mu} \left( \dot{r} \,f_r + r \,\dot{\nu}\, f_\nu \right) \,,\\
	& \dot{e} = \frac{1 - e^2}{e} \left(\frac{\dot{a}}{2\,a} - \frac{\dot{h}}{h} \right) \,,\\
	& \dot{\omega} = \frac{h}{e \mu} \left( \frac{2+e\cos{\nu}}{1+e\cos{\nu}} f_\nu \sin{\nu} - \cos{\nu} f_r \right) \label{eq:perturbeq3}\,,\\
	&\dot{\nu} = \frac{(1+e\cos{\nu})^2}{(1-e^2)^{3/2}} \sqrt{\frac{\mu}{a^3}} - \dot{\omega}\,,
\end{align}
where $h = |\mathbf{r} \times \mathbf{v}|$ is the magnitude of the specific angular momentum \REV{and} $\mu = G (m_1 + m_2)$ is the standard gravitational parameter.

The above equations describe the evolution of the binary as a function of time. However, they are still phase-dependent, in the sense that they depend on the true anomaly $\nu$ of the binary at any given time.
On the other hand, the equations that are commonly employed in BPS codes are orbit-averaged.

We can derive the secular equations from Equations~\ref{eq:perturbeq}--\ref{eq:perturbeq3} by averaging over the mean anomaly $M$.
Given a phase-dependent derivative $\dot{q}(\nu)$, the corresponding orbit-averaged equations can be derived as:
\begin{equation}\label{eq:orbitaveraged}
	\langle \dot{q} \rangle  = \frac{1}{2\pi} \int_0^{2\pi} \dot{q}(\nu) dM = \frac{1}{2\pi} \int_0^{2\pi} \frac{(1-e^2)^{3/2}}{(1+e\cos{\nu})^2}  \dot{q}(\nu) d\nu\,.
\end{equation}

In the next sections, we focus mainly on the phase-dependent equations.
The orbit-averaged expressions suitable for the inclusion in BPS codes are presented in the Appendix~\ref{sec:secular}.

\subsection{Power-law radial dependence}\label{sec:forcepowlaw}

A general form for $P$ in \autoref{eq:dragform} is a power-law $P(r) = r^{-k}$.
In this case, the acceleration has a magnitude of
\begin{equation}\label{eq:forcepowlaw}
	f = - C \frac{v^l}{r^k}\,,
\end{equation}
and $C$ has physical dimensions $[C] = L^{1-l+k} T^{l-2}$. 
We choose a power-law force mainly because it is analytically convenient to treat.
In our forthcoming work, we will focus on a more realistic density profile for the envelope.

Keeping the exponents $l$ and $k$, we derive the following set of ordinary differential equations for the orbital elements:
\begin{widetext}
\begin{align}\label{eq:powerlawkl}
	\dot{a} & = - 2 C \, \mu^{\frac{l-1}{2}} \, a^{\frac{3-l-2k}{2}} \, (1-e^2)^{-\frac{l+1+2k}{2}} \, (1+e\cos{\nu})^k \, (1+e^2 + 2e\cos{\nu})^{\frac{l+1}{2}} \,,\\\label{eq:powerlawkl2}
	\dot{e} & = - 2 C \, \mu^{\frac{l-1}{2}} \, a^{\frac{1-l-2k}{2}} \, (1-e^2)^{-\frac{l-1+2k}{2}} \, (1 + e \cos{\nu})^k \, (1+e^2 + 2 e \cos{\nu})^{\frac{l-1}{2}} \,\left(e + \cos{\nu}\right) \,,\\\label{eq:powerlawkl3}
	\dot{\omega} & = - 2 C \mu^{\frac{l-1}{2}}\, a^{-\frac{l-1+2k}{2}} \,\frac{(1-e^2)^{-\frac{l-1+2k}{2}}}{e}\, (1+e\cos{\nu})^k \, (1+e^2 + 2 e \cos{\nu})^{\frac{l-1}{2}} \, \sin{\nu} \,,\\\label{eq:powerlawkl4}
	\dot{\nu} & = \frac{(1+e\cos{\nu})^2}{(1-e^2)^{3/2}} \sqrt{\frac{\mu}{a^3}} - \dot{\omega}\,.
\end{align}
\end{widetext}

For $l=2$ (and even numbers), Equations~\ref{eq:powerlawkl}--\ref{eq:powerlawkl4} gain a term in $\sqrt{1+e^2 + 2 e \cos{\nu}}$ which makes the integral in the orbit-averaged \autoref{eq:orbitaveraged} impossible to be expressed in closed form because it gives rise to an elliptic integral.
Alternatively, the elliptic integral can be tabulated as a function of $e$.

The value of $C$ sets the timescale of the CE inspiral, and it relates to the density of gas in the envelope.
However, we treat it at first as a given constant to investigate the qualitative behavior of Equations~\ref{eq:powerlawkl}--\ref{eq:powerlawkl4}.
We therefore define a dimensionless CE efficiency $\chi_a = P / \tau_{a}$, where $P$ is the binary period and $\tau_{a} = |a / \dot{a}|$ is the characteristic timescale of the binary's inspiral.
Setting $e=0$ in \autoref{eq:powerlawkl}, the expression for $\chi_a$ is:
\begin{equation}\label{eq:chia}
	\chi_a = C \pi \mu^{\frac{l-2}{2}} a^{\frac{4-l-2k}{2}}\,.
\end{equation}

We can then use \autoref{eq:chia} to find the values of $C$ for models with different $k,l$ but similar inspiral time.
For the perturbative equations to be valid, the drag force needs to be weaker than the mutual gravitational acceleration, hence $0 < \chi_a \ll 1$. 

We now integrate numerically Equations~\ref{eq:powerlawkl}--\ref{eq:powerlawkl4} and compare them with direct $N$-body integration, in which we apply the force of \autoref{eq:forcepowlaw} directly as a perturbative force over the Newtonian equations.

For the integration of the perturbative equations, we employ an adaptive Runge-Kutta method of order 8(5,3).
The direct $N$-body integration is computed with the Hermite integrator from the \textsc{amuse} software environment \citep{hut1995,amusebook}.
The drag force is added as a velocity kick every 1/100th of an orbital period.

In the following examples, the binary star has masses $m_1 = 81 \msun$ and $m_2 = 32 \msun$, an initial semimajor axis $a_0 = 4000 \rsun$ and eccentricity $e_0 = 0.2$.
Because we have not yet introduced a self-limiting mechanism to stop the inspiral, we stop the integration once the binary semimajor axis reaches $40 \rsun$.
For both the $N$-body and the perturbative integration we choose $l=1,2, k=0$, and set $\chi_a = 0.05$.
The binary is initialized just before pericenter passage, at $\nu = 270^\circ$. The initial argument of pericenter is $\omega = 90^\circ$.

\begin{figure} 
	\centering
	\includegraphics[width=1\linewidth,trim=0.35cm 0.25cm 0.38cm 0.38cm,clip]{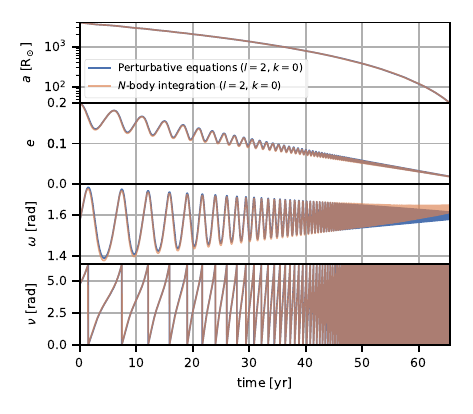}
	\includegraphics[width=1\linewidth,trim=0.35cm 0.25cm 0.38cm 0.38cm,clip]{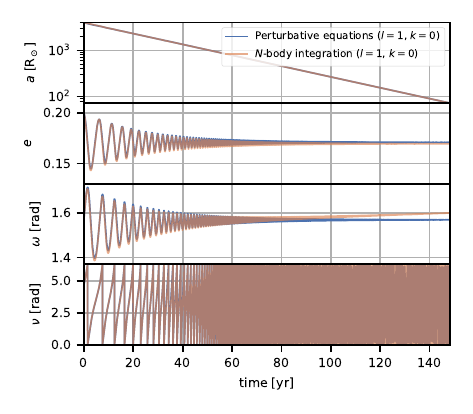}
	\caption{Evolution of semimajor axis ($a$), eccentricity $e$, argument of pericenter $\omega$ and true anomaly $\nu$ for a CE inspiral using the model in \autoref{eq:forcepowlaw} with $l=2,k=0$ (top panel) and $l=1,k=0$ (bottom panel). The blue curves were obtained integrating Equations~\ref{eq:powerlawkl}--\ref{eq:powerlawkl4}, while the orange curves are the result of direct $N$-body integration with the drag force applied as a perturbation. The two curves overlap to a great extent.}
	\label{fig:nbodycomp}
\end{figure}

Figure~\ref{fig:nbodycomp} shows the result of the integrations for $l=2$ and $l=1$.
The two curves agree to a satisfactory degree, but the perturbative equations take ${\sim}50$ less computational time to integrate.
For $l=2$, the binary eccentricity decreases on average but oscillates during one orbital period, with the eccentricity decreasing at the pericenter and increasing at the apocenter.
In contrast, for $l=1$, the eccentricity oscillates around a constant value and the inspiral proceeds at constant eccentricity. 
In both cases, the long-term evolution of $\omega$ appears to disagree between the perturbative equation and direct integration.
As apparent from \autoref{eq:powerlawkl3}, the argument of pericenter is supposed to only oscillate around a median value, without any long-term deviation.
We attribute the discrepancy to the simple implementation of our $N$-body model rather than to some inaccuracy of the perturbative equations. \REV{Performing the orbit-averaging on Equations~\ref{eq:powerlawkl2}--\ref{eq:powerlawkl3} for $l=1, k=0$, further confirms that the eccentricity and argument of pericenter remain constant over long timescales (see Appendix~\ref{sec:massloss}).}

It is clear that a drag force linear in the velocity ($l=1$) does not agree with our naive expectations of CE evolution circularizing the binary.
We will see later that introducing a radial dependency in the force can quickly circularize the binary, even for $l=1$.
Having now validated our set of equations with the $N$-body integration, we now investigate the effect of radial dependency, controlled by the parameter $k$.

\begin{figure}
	\centering
	\includegraphics[width=1\linewidth,trim=0.35cm 0.25cm 0.38cm 0.35cm,clip]{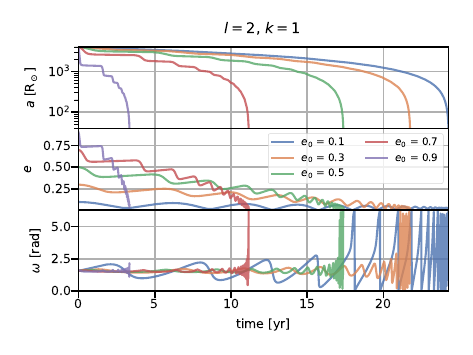}
	\includegraphics[width=1\linewidth,trim=0.35cm 0.25cm 0.38cm 0.35cm,clip]{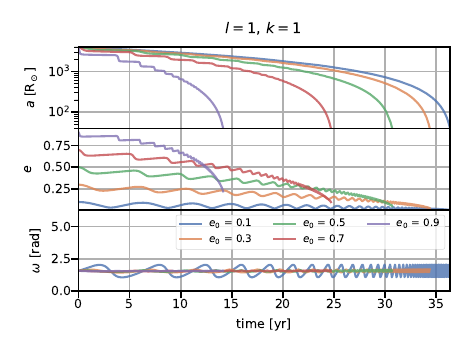}
	\includegraphics[width=1\linewidth,trim=0.35cm 0.25cm 0.38cm 0.35cm,clip]{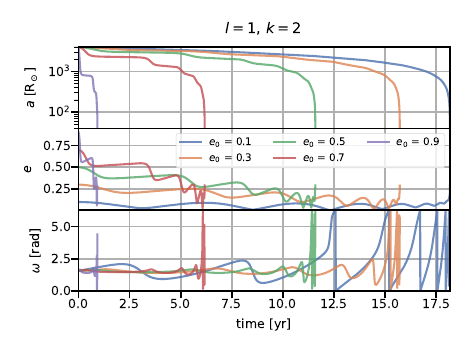}
	\caption{Evolution of semimajor axis $a$, eccentricity $e$, argument of pericenter $\omega$. For a CE inspiral using the model in \autoref{eq:forcepowlaw} with $l=2,k=1$ (top panel), $l=1,k=2$ (middle panel) and $l=1,k=2$ (bottom panel). Each curve indicates a different starting eccentricity.}
	\label{fig:kcomp}
\end{figure}

The same system integrated with $k=1$ and $k=2$ is shown in \autoref{fig:kcomp}.
Here we set the binary with different initial eccentricities to appreciate the effect of radial dependency on the circularization timescale.
Even a moderate radial dependency ($k=1, f \propto 1/r$) introduces strong orbit circularization, and the drop in the semimajor axis occurs mostly at the pericenter.

For $l=2, k=1$, the final eccentricity after the binary reaches our target semimajor axis is negligible, while for $l=1, k=1$ the binary still retains some eccentricity after the inspiral.

\subsection{Drag force from first principles}

Let us now examine how the expression for the force should depend on $r$ and $v$ using only theoretical principles.
Besides the explicit dependency in $v^2$, \autoref{eq:dragphys} may include a hidden dependency in the cross-section area $A$.
In our case, the cross-section identifies the size of the sphere around the perturbing body where the gas becomes gravitationally focused.
According to the Hoyle-Lyttleton-Bondi accretion theory \citep{hoyle1939,bondi1944,bondi1952}, the radius of this sphere is
\begin{equation}\label{eq:accrad}
	R_{\rm a} = \frac{2 G m }{c_{\rm s}^2 + v^2}\,,
\end{equation}
where $c_{\rm s}$ is the sound speed of the gas and $m$ is the mass of the perturber.
Given that the orbital velocity is much higher than the sound speed in the envelope, we can approximate $R_{\rm a} \approx 2 G m / v^2$.
Consequently, $A \propto v^{-4}$ and the force appears to scale inversely proportional with the velocity: $f \propto v^{-2}$. 
To first order, the scaling $v^{-2}$ is the same as that of the dynamical friction force in collisionless or gaseous media \citep{chandrasekhar1943,ostriker1999,kim2007,kim2008,muto2011,grishin2015,grishin2016}.
\REV{However, integrating \autoref{eq:powerlawkl2} reveals that for negative $l$ the eccentricity \textit{increases} rather than decreases with time. The physical reason is the following.
In the Hoyle-Lyttleton-Bondi accretion model, as the velocity increases, the size of the gravitationally focused fluid decreases accordingly, leading to the drag force being stronger at the apocenter, where the orbital velocity has a minimum, rather than at the pericenter, where the velocity is the highest. 
Consequently, the perturber loses more energy and angular momentum at apocenter than at pericenter. The net effect is that the orbit becomes more radial after the apocenter passage than it can circularize at the pericenter.	

This is especially true for $l=-2$. Large values of $k$ can mitigate the increase of the eccentricity, because the factor $r^{-k}$ makes the drag force stronger at pericenter than at apocenter. However, only a very steep radial density profile can prevent the growth of the eccentricity for $l=-2$. Numerical integration shows that the eccentricity grows faster than exponentially for $k<3$. In practice, for $l=-2, k<3$, a single apocenter passage at high eccentricities ($e \gtrsim 0.5$) can make the orbit radial $e\simeq 1$. We demonstrate this rigorously in the Appendix~\ref{sec:lnegative2}, where we analyze the evolution of the eccentricity for $l=-2$ by orbit-averaging \autoref{eq:powerlawkl2}. This behavior is clearly an artifact of the simplified physics adopted in the dynamical friction model. One issue can stem from the fact that, as the eccentricity increases, the ram pressure at the pericenter also increases, likely supplanting the gravitational drag as the force driving the orbital decay.
Moreover, dynamical friction forces with scaling ${\propto} v^{-2}$ are based on the assumption of a body traveling in a uniform, infinite medium, while the envelope of a giant producing the drag force on companion is not uniform nor infinite. The gravitationally focused fluid does not travel in a straight line, always directly behind the companion, but follows perturbed Keplerian trajectories. Finally, spiral density waves that generate during the inspiral can also apply a torque on the binary, but their effect cannot be easily modeled without introducing axisymmetry.}
For these reasons, hereafter we consider only positive values of $l$ of the drag force model, considering it an `effective' drag force rather than one purely caused by dynamical friction and gravitational focusing.
Our choice is further confirmed by the comparison with hydrodynamical simulations (\autoref{sec:hydrocomp}), which consistently show a decrease in the eccentricity \citep{glanz2021b}, and previous numerical studies on the gravitational drag force in wind tunnel simulations \cite{shima1985,macleod2017b}.

\REV{In general, the loss of tangential velocity due to the dynamical friction is necessary for the inspiral motion at the first place, and acts against the tidal circularization. In addition, past studies and simulations shows that even an initial circular orbit developed some eccentricity by the end of the inspiral \citep{soker2000,passy2012,staff2016,kashi2018,glanz2021b}. }

On the other hand, values of $k$ may be linked to the radial density profile of the giant star.
Whether or not the envelope density profile could be approximated by a power-law, one-dimensional stellar profile is unlikely to represent the stellar structure during the CE inspiral because the presence of the secondary into the primary's envelope will significantly affect its density profile.
Besides this, the stellar structure of the primary might be altered even prior to the inspiral phase due to mass transfer and tidal forces. 
These considerations further motivate us to regard the profile in \autoref{eq:forcepowlaw} as an effective force rather than strictly associate it to a specific physical mechanism.

\subsection{Analytic solutions for zero eccentricity}

For zero initial eccentricity, the dependency on the true anomaly disappears, and we are left with a single differential equation for $a$:
\begin{equation}\label{eq:zeroeccadot}
	\dot{a} = - 2 C \mu^\frac{l-1}{2} a^\frac{3-l-2k}{2}\,.
\end{equation}
Defining the parameter $m= (3-l-2k)/2$, this equation has the following solutions:
\begin{align}\label{eq:analsol}
		&a = \sqrt[1-m]{a_0^{1-m} - 2(1-m) C t \mu^\frac{l-1}{2}} &\quad m\neq1\,,\\
		&a = a_0 e^{-2Ct\mu^{(l-1)/2}} & \quad m=1\,.
\end{align}

The decay is exponential for $m=1$, which corresponds to the case $l=1$, $k=0$.

\section{Halting the inspiral}\label{sec:halt}

So far, we have neglected any self-limiting mechanism to the binary inspiral.
However, in a realistic CE evolution, the orbital energy is spent on unbinding the envelope, which may result in its complete ejection. 
The detailed mechanism of envelope ejection during a CE event is still under discussion.
In many hydrodynamical simulations, the envelope is never fully ejected, but it simply extends to larger separations while remaining bound to the binary \citep{ricker2008,demarco2011,ricker2012,passy2012,nandez2015,ohlmann2016,staff2016}.
Additional mechanisms, such as hydrogen and helium recombination \citep{ivanova2015,ivanova2016,reichardt2020,lau2022a,lau2022b} or dust-driven winds \cite{glanz2018}, have been invoked to explain the complete envelope ejection. 

While our simple 0-dimensional model cannot capture as many details as a full 3D hydrodynamical simulation, it is still tempting to seek a self-limiting mechanism for our CE model, which may be analytically tractable.
In this section, we discuss the possible models and their limitations.

\subsection{Using the $\alpha \lambda$ formalism}\label{sec:alphalambdaconsist}

The simplest way to halt the inspiral phase is to rely on the $\alpha \lambda$ model. 
In the $\alpha \lambda$, the orbital energy lost during CE is calculated on the basis of an energy balance equation.
The binding energy of the envelope is parametrized by $\lambda$ in the following expression \citep{dekool1990}:
\begin{equation}\label{eq:bindenergylambda}
	E_{\rm bind} = -\frac{G m_{1\rm, env} m_{1}}{\lambda R}\,,
\end{equation}
where $m_{1\rm, env}$ is the mass of the envelope.
When both stars are giants at the onset of CE, the binding energy of both envelopes is included in \autoref{eq:bindenergylambda}.

The binding energy is compared to the difference in orbital energy
\begin{equation}\label{eq:orbenergy}
	\Delta E_{\rm orb} = -\frac{G m_{1\rm, c} m_2}{2a_{\rm f}} + \frac{G m_1 m_2}{2a_{\rm i}}\,,
\end{equation}
where $a_{\rm i}$ and $a_{\rm f}$ are the pre- and post-CE semimajor axes.

The parameter $\alpha$ is introduced before equating \autoref{eq:bindenergylambda} and \autoref{eq:orbenergy}, and represents the CE efficiency.
The final expression from which the final semimajor axis $a_f$ is calculated reads as:
\begin{equation}\label{eq:alphalambda}
	\frac{G m_{1\rm, c} m_{1\rm, env}}{R} = \alpha \lambda \left(\frac{G m_{1\rm, c} m_2}{2a_{\rm f}} - \frac{G m_1 m_2}{2a_{\rm i}} \right)	\,.
\end{equation}
In BPS codes, sometimes $\alpha \lambda$ are used together as a single parameter, although this approach creates a degeneracy between the binding energy estimate and CE efficiency.
A better approach is to estimate $\lambda$ from detailed 1D stellar evolution models \cite{hurley2002,dewi2000,xu2010,xu2010erratum,claeys2014,kruckow2018}. 

We can use the $\alpha \lambda$ formalism together with the drag force formalism presented in \autoref{sec:dragforceform}.
This choice has the advantage of producing results consistent with the $\alpha \lambda$ (by construction), while still allowing for non-zero final eccentricity and avoiding discontinuities in the evolution of the semimajor axis.

We can express the expected orbital energy loss by rewriting \autoref{eq:alphalambda} as
\begin{equation}\label{eq:deltaeorb}
	\Delta E_{\rm orb} = \frac{1}{\alpha \lambda} \frac{G m_{1} m_{1\rm, env}}{R}\,.
\end{equation}
While using the drag force formalism, we can then keep track of the orbital energy loss as
\begin{equation}\label{eq:enelossaonly}
	\dot{E}_{\rm orb} = \frac{G m_1 m_2}{2 a^2} \dot{a}\,,
\end{equation}
and we can stop the integration as soon as the accumulated orbital energy loss equates \autoref{eq:deltaeorb}:
\begin{equation}\label{eq:integratealphalambda}
	\Delta E_{\rm orb} = \int_{t_0}^{t_\mathrm{end}} \dot{E}_{\rm orb} dt\,.
\end{equation} 

As stated earlier, \autoref{eq:enelossaonly} takes into account for the energy losses caused by drag force only.
Although it is straightforward to add the additional derivative terms accounting for mass changes (see Appendix~\ref{sec:massloss}), finding an expression for the mass loss $\dot{M}:=\dot{m}_\mathrm{env}$ consistent with the $\alpha \lambda$ model is not as simple.
The reason is that the $\alpha \lambda$ model is fine-tuned to result in the complete ejection of the envelope as soon as the inspiral ends.
This is possible because it uses a simple energy balance equation. However, the drag force model decouples mass loss and energy loss, allowing for inspirals to end before the envelope is fully ejected.
A possible way to couple mass loss and inspiral is to impose that the mass loss $\dot{m}_\mathrm{env}$ is proportional to the rate of orbital decay due to the drag force, times the remaining envelope mass, $\dot{m}_\mathrm{env} \propto m_\mathrm{env} \dot{a}/a$.
However, this coupling would need to be precisely fine-tuned to achieve the envelope ejection only when the total $\Delta E_{\rm orb}$ equates to the amount prescribed by the $\alpha \lambda$ model. 

\REV{We believe that this is not a drawback of the model, but a feature that enables a more realistic modeling of the CE phase. In fact, recent studies point out that a fraction of the envelope will likely remain bound to the core after the inspiral has stalled \citep{fragos2019,vignagomez2022}. }

\subsection{Self-similar expansion}\label{sec:selfexp}

\begin{figure}[b]
	\centering
	\includegraphics[width=1\linewidth,trim=0.38cm 0.35cm 0.38cm 0.4cm,clip]{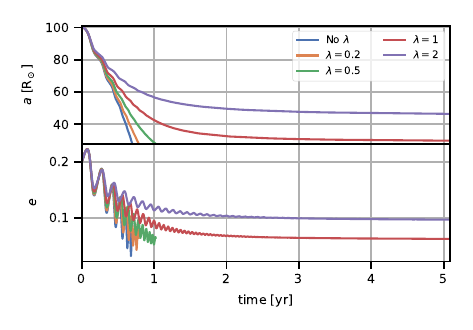}
	\caption{Evolution of semimajor axis (top) and eccentricity (bottom) as a function of time, assuming self-similar expansion of the envelope. The colors indicate different initial binding energy of the envelope, parametrized by the $\lambda$ parameter (\autoref{eq:bindenergylambda}). Blue: no envelope expansion. Orange: $\lambda=0.2$. Green: $\lambda=0.5$. Red: $\lambda=1$. Purple: $\lambda=2$. We set $\chi_a = 0.05$. }
	\label{fig:expansion}
\end{figure}

In reality, the energy loss goes into unbinding the envelope, so that its density decreases, the drag force becomes weaker, and the inspiral stalls.

The binding energy of a non-rotating star can be calculated with the following integral: 
\begin{equation}\label{eq:bindenergy}
 E_{\rm bind} = \int^{M}_{m{_{\rm c}}}(E_{\rm int} - \frac{Gm}{r})dm\,,
\end{equation}
where $E_{\rm int}$ is the internal energy, while the second term is the gravitational term.
In the past, only the gravitational binding energy was taken into account, but recent works have begun to include the thermal energy $E_{\rm int}$ term and even recombination energy \citep{claeys2014,kruckow2016,xu2010,klencki2021}. 

A simple approach is to assume that the envelope expands homologously, conserving mass.
Given an expansion factor $g(t)$, the density $\rho$ as a function of time and position can be expressed as:
\begin{equation}\label{eq:densityexp}
	\rho(t, r) = \frac{1}{g(t)^3} \rho_0 \left(\frac{r}{g(t)}\right) \,,
\end{equation}
where $\rho_0$ is the original density at time $t=0$ so that $g(t=0) = 1$ (see \citep{ginat2020} for an analogous method with a constant expansion factor).
Intuitively, the density at position $r$ and time $t$ is the density of the original profile at the old position $r' = r / g$, rescaled by a factor $g^3$ in order to conserve the mass.
The radius of the envelope at time $t$ is therefore $R(t) = R_0 g(t)$.
For a polytropic sphere, the binding energy $B$ is proportional to the inverse of the radius, so that we can write:

\begin{equation}
	B(t) = \frac{B_0}{g(t)} \,.
\end{equation}

Equating the orbital energy losses $\dot{E}_{\rm orb}$ to the binding energy losses:
\begin{equation}
	\dot{B} = - \dot{g} \frac{B_0}{g^2} = \dot{E}_{\rm orb}\,,
\end{equation}
so that the expansion factor $g(t)$ evolves as:
\begin{equation}
	\dot{g} = - \frac{m_{\rm red}\mu}{2a^2} \frac{\dot{a}}{B_0} g^2 \,,
\end{equation}
where $m_{\rm red}$ is the reduced mass.
This equation can be integrated alongside Equations~\ref{eq:powerlawkl}--\ref{eq:powerlawkl4}, but requires an initial estimate of the initial binding energy $B_0$.
It is possible to use the classic $\lambda$ parametrization of $B_0$ (\autoref{eq:bindenergylambda}).

Changes in the local density $\rho(t,r)$ reflect on the drag force through the parameter $C$.
Because the power-law model assumes that $\rho_0(r) \propto r^{-k}$, it follows from \autoref{eq:densityexp} that $C(t) = C_0 / g(t)^{3-k}$, which closes our set of equations.

Figure~\ref{fig:expansion} shows the common envelope evolution of a $1 \msun$ giant, $0.6\msun$ companion binary with the inclusion of the expansion factor.
Here the initial orbit has an eccentricity of 0.2 and a semimajor axis of $100 \au$.
To estimate the initial binding energy, we assume that the radius of the giant star is $R=83 \rsun$ and its core mass $0.39 \msun$.
The introduction of the envelope expansion makes the inspiral self-limiting: as the gas density decreases, the inspiral slows down until it stalls.
The final value of both eccentricity and semimajor axis depends on the value of the binding energy, parametrized by $\lambda$.
Lower binding energies (larger values of $\lambda$) halt the inspiral sooner, leaving the binary at larger separations and higher eccentricity.




\section{Comparison with hydrodynamical simulations}\label{sec:hydrocomp}

\begin{figure*}
	\centering
	\includegraphics[width=1\linewidth,trim=0.38cm 0.42cm 0.42cm 0.38cm,clip]{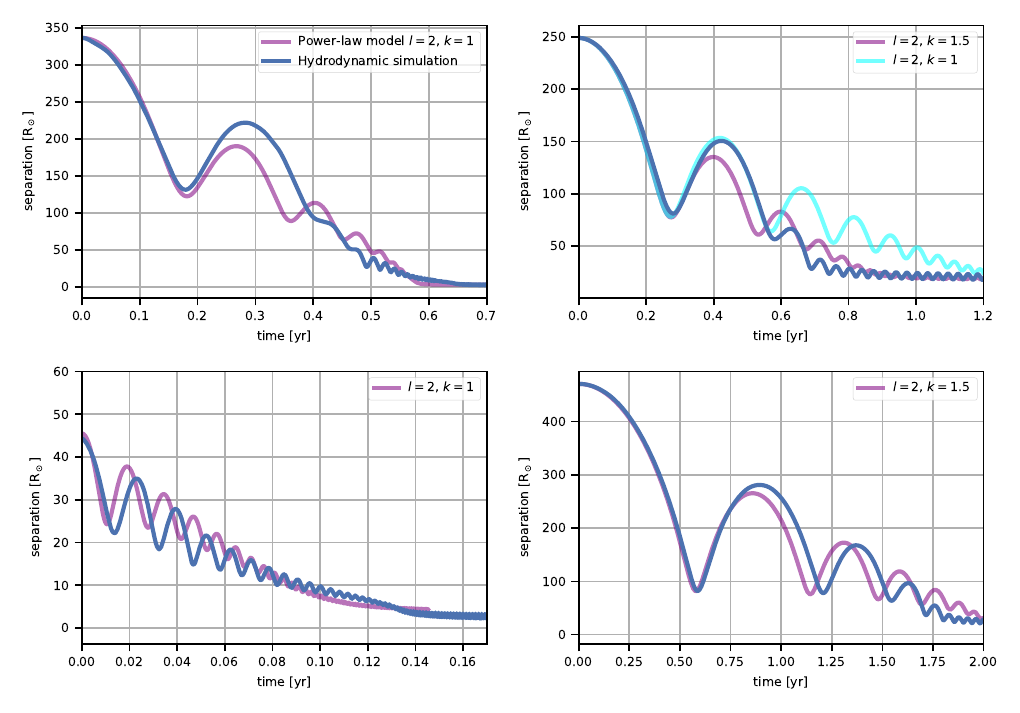}
	\caption{Comparison of the binary separation evolution for four hydrodynamical simulations and the power-law drag force model. From top to bottom, left to right: simulations \textsc{8R2G5}, \textsc{8R2G-0}, \textsc{1R06P5}, \textsc{1R06P7} from \citet{glanz2021b}. For clarity, we compared models in \textsc{8R2G-0} from the second apocenter approach, where the CE rapid plunge-in began. The blue line is the separation obtained from the hydrodynamical simulations, while the thick purple line is the power-law model. All the semi-analytic curves use the quadratic drag force ($l=2$). The left panels use a drag force linearly decreasing with radius ($k=1$), while the right panel adopts a power-law of $k=3/2$. The top-right panel shows the difference between $k=1$ and $k=3/2$.}
	\label{fig:hydrocomp}
\end{figure*}

As explained in \autoref{sec:forcepowlaw}, the drag force model can be considered as an effective model that can reproduce the underlying physics of common envelope inspiral.
It is difficult to assess a priori which values of $l$, $k$, and $C$ better describe the complex physics at play.
For this reason, we compare the power-law model with smoothed-particle hydrodynamical simulations of eccentric CE run by \citep{glanz2021b}.
\autoref{fig:hydrocomp} compares the core separation between the semi-analytic model and four simulations with different stellar parameters and initial orbital eccentricities. We note, that while here we consider the location of the cores where the center of mass of each star is, this is not necessarily the case in the hydrodynamical simulations, where the envelope is not limited to a spherical expansion.
The left panels of \autoref{fig:hydrocomp} show a CE event between a $8 \msun$ giant and a $2 \msun$, while the right panels involve a $1 \msun$ giant with a $0.6 \msun$ companion.
We compare the hydrodynamical simulations with the self-similar expansion model of \autoref{sec:selfexp}, choosing different values of $k$, $C$ and $\lambda$, but fixed $l=2$. 

The semi-analytic model can reproduce the CE inspiral of the hydro simulations with relatively good agreement.
\REV{
Even though the 1D density profile of the giant star's envelope roughly follows a power-law exponent of ${\lesssim}3$, the inspiral is well reproduced by $k=1$.
For the right panels of \autoref{fig:hydrocomp}, the simulations with $ 1 \msun$ giant and a $0.6 \msun$ companion have a much faster inspiral than the $k=1$ model, and $k=1.5$ better matches the separation decay.

The two different stellar models are better matched by a different power-law index $k$, confirming our ansatz that the radial part $P(r)$ of the drag force represents the local density. On the other hand, the estimated power-law $k$ is significantly shallower than the radial density profile of the 1D stellar evolution model. This might be caused by the fact that the stellar profile has been significantly altered by the tidal field of the companion. In real systems, the stellar profile might have been further altered by mass transfer and tidal spin up.

In addition to the stellar density profile, the equation of state of the star plays a crucial role in the unbinding of the CE material and the stalling of the inspiral. In particular, more massive stars are radiation-pressure dominated, and therefore they have a lower binding energy than low-mass stars. Hydrodynamics simulations have showed that CE events of red supergiant stars end at a larger separation and with a higher fraction of unbound mass when radiation pressure and recombination energy contributions are included \citep[][see also \citealp{klencki2021}]{lau2022a,lau2022b}.

In our model, besides adopting a different value of $k$, differences between massive and low-mass stars can be incorporated into the initial binding energy $B_0$ when using the self-similar expansion assumption, or into the parameter $\lambda$, in the same way it is to the $\alpha\lambda$ model. This is agreement with the best-match values of $\lambda$ estimated from the comparison in \autoref{fig:hydrocomp}, which are $\lambda \simeq 0.25$ for the low-mass model and $\lambda \simeq 0.5$ for the high-mass one. 

}

\section{Applications to astrophysical scenarios}

\subsection{CE triggered by the von~Zeipel-Kozai-Lidov mechanism in triples}\label{sec:CEkozai}

In hierarchical triple systems, a stellar binary is orbited by a tertiary star.
The tertiary may affect the orbital evolution of the binary through gravitational interactions, giving rise to the so-called von~Zeipel-Kozai-Lidov (ZKL) mechanism, wherein the eccentricity of the inner binary can be excited to extremely high values \citep{zeipel1910,lid62,koz62,naoz2016,shev2017}.
The ZKL mechanism has been invoked to explain a variety of astrophysical phenomena, for example, gravitational wave mergers of compact objects \citep[e.g.][]{trani2022}, type Ia supernovae detonation from white dwarf collisions \citep[e.g.][]{kushnir2013}, or evolutionary pathways in interacting stellar triples \citep[e.g.][]{toonen2020}.

\begin{figure}
	\centering
	\includegraphics[width=1\linewidth,trim=0.35cm 0.40cm 0.40cm 0.38cm,clip]{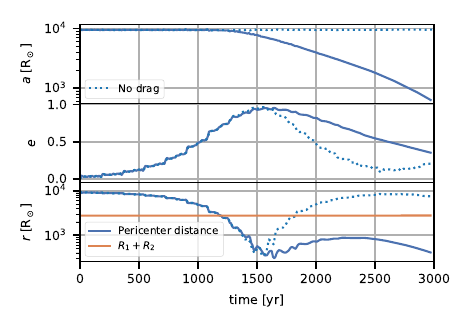}
	\caption{Semimajor axis (top), eccentricity (middle), and separation (bottom) of the inner binary of a triple stellar system as a function of time. Solid lines: simulation including a drag force term with $l,k=2,0$ and $\chi_a = 0.005$. Dotted line: simulation with no drag force. We stop the CE simulation as soon as the energy released by the drag equals the binding energy of the envelope. Evolved by means of direct $N$-body integration with an Hermite scheme \citep{hut1995,amusebook}.}
	\label{fig:triplevol}
\end{figure}

\begin{figure*}
	\centering
	\includegraphics[width=1\linewidth,trim=0.38cm 0.42cm 0.40cm 0.38cm,clip]{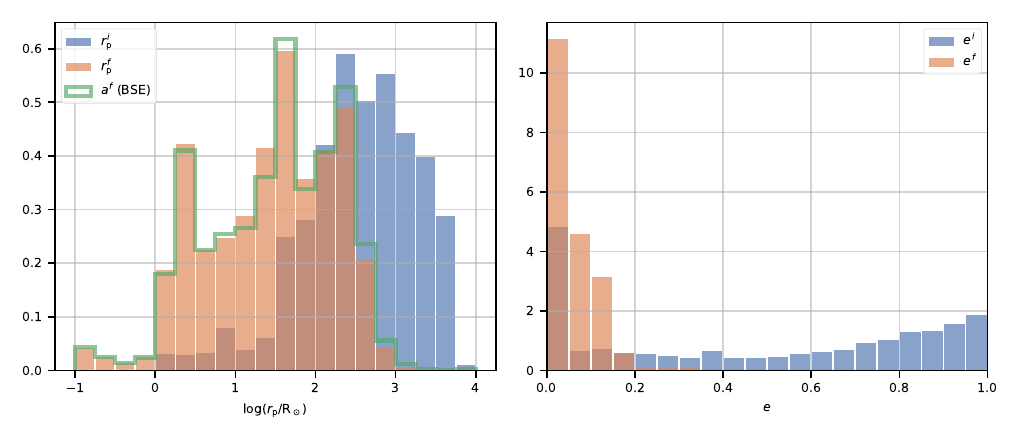}
	\caption{Pericenter distance (left) and eccentricity (right) distributions, pre-CE (blue) and post-CE (red). \REV{In the left panel, the post-CE semimajor axes distribution derived from \textsc{bse} is outlined in green.} Using the drag force model with $\chi_a=0.05$, $l=2$, $k=1$, and stopping the inspiral to be consistent with the $\alpha \lambda$ model, as described in Section~\ref{sec:alphalambdaconsist}. The common envelope events are taken from the BPS calculations of \citet{tanikawa2020b}. In BPS codes, the final eccentricities are in practice always zero, while we obtain small but finite eccentricities in our model.}
	\label{fig:alphalambda}
\end{figure*}

In stellar triples, the high eccentricity during a ZKL cycle can trigger tidal interaction, mass transfer, and even CE evolution \citep[e.g.][]{toonen2020,grishin2022}.
While it is straightforward to include tidal and mass transfer interactions in $N$-body or triple stellar evolution codes \citep{toonen2016,trani2020b,hamers2021b, hamers2022}, no analytic self-consistent CE model for few-body systems exists yet \citep[see also][]{comerford2020,glanz2021a}. 
Here we show how our drag force model can be included in direct $N$-body codes, enabling the modeling of CE in hierarchical triples and even higher multiple systems.

\autoref{fig:triplevol} shows the evolution of a triple system consisting of a $66 \msun$, $2771 \rsun$ giant with a $29 \msun$ main sequence companion, and a tertiary $29 \msun$ star.
The tertiary is inclined by 90$^\circ$ with respect to the inner binary, which gives rise to ZKL oscillations.
\REV{During the first ZLK cycle, the eccentricity of the inner binary grows enough that the secondary enters into the envelope of the primary.
In the absence of drag forces or collisions, the ZLK cycle continues and the eccentricity naturally decreases again (dotted lines in \autoref{fig:triplevol}). This happens because, in the absence of binary interactions, this configuration is stable and the triple could undergo another ZLK cycle, with the eccentricity oscillating between ${\sim}0.06$ and $0.96$.}
When we include the drag force, the binary semimajor axis begins to decay as soon as the secondary is engulfed by the primary (solid lines in \autoref{fig:triplevol}).
However, the drag is too weak to damp the increase in eccentricity, which continues to rise and decay following the ZKL cycle.
However, as the semimajor axis decreases, the perturbation from the tertiary star becomes weaker, and the eccentricity decreases slowly.
We stop the simulation once the energy released by the drag force matches the binding energy of the envelope, as calculated using the prescriptions of \citet{claeys2014}.
In the end, the semimajor axis has shrunk by a factor of 20, retaining an eccentricity of ${\sim} 0.35$. 
Conversely, $\alpha\lambda$ model predicts a similar final semimajor axis ($a_{\rm f}\simeq 634 \rsun$), but with zero final eccentricity.

In this example, we focused on the main differences between our common envelope prescription and the $\alpha\lambda$ model.
Therefore, we have neglected other forces, like tidal interactions or mass transfers prior to the CE event. 
Additional forces to model these effects can be easily included \citep{hut81,dosopoulou2016}, but they would be limited to altering the binary parameters prior to the CE event.

\subsection{CE in isolated binary stellar evolution}\label{sec:bspcomparison}

BPS codes employ the $\alpha\lambda$ model, meaning that they cannot estimate the eccentricity of binaries after CE events.
In the \textsc{bse} code and derivatives, the final eccentricity is set to zero after practically every CE evolution. 
However, this is in tension with the observation of systems believed to be post-CE binaries, such as close binaries with a subdwarf B/O star.
In fact, while most studies assume zero eccentricity for such systems, a few post-CE binaries show eccentricities up to ${\sim}0.15$ \citep{delfosse1999,edelmann2005,lynch2012,kawka2015,kruckow2021}.

We have extracted a sample of CE episodes from 2659 runs of a modified version of \textsc{bse} \citep{tanikawa2020b}, and we have re-run them with our new model, using the perturbative equations of \autoref{sec:forcepowlaw}.
The \textsc{bse} runs were obtained from the following setup.
The initial mass function (IMF) of primary stars follows Kroupa's IMF \citep{kroupa2001} with the minimum and maximum masses of $10$ and $150\, \msun$, respectively.
Mass ratios of secondary stars to primary stars are distributed uniformly between 0 and 1.
The semi-major axis distribution is uniform in a logarithmic scale between 1 and $10^6 \rsun$.
The eccentricity distribution is thermal.
The single star evolution is Hurley's model \citep{hurley2000} with stellar metallicity $Z=0.002$.
We adopt common envelope parameters $\alpha=1$ and Claeys's $\lambda$ \citep{claeys2014}.
The common envelope evolution sets in under the same criteria as \cite{hurley2002}. 

\REV{
\textsc{bse} already includes recipes for the tidal circularization of close binaries and orbital changes due to mass and momentum transfer. Nonetheless, many systems enter the CE phase with a significant eccentricity, as shown by the blue distributions in \autoref{fig:alphalambda}. This is consistent with the population synthesis study of \citet{vignagomez2020}, which estimated that at least $18\%$ of the binaries will be eccentric at the onset of the Roche lobe overflow that leads to the CE event. This result is also supported by detailed studies on the tidal evolution of evolving giants with close companions \citep{vick2021}.}

For all the runs with our common envelope formalism, we adopt $l=2$, $k=1$ and $\chi_a = 0.05$. As described in \autoref{sec:alphalambdaconsist}, we make our model consistent with the $\alpha\lambda$ model by stopping the integration once the energy released by the drag force equals the binding energy of the envelope. We also avoid applying the drag once the secondary is outside the envelope of the primary. 

Our model allows to estimate the eccentricities after CE, which are not necessarily zero. This is clear from
\autoref{fig:alphalambda}, which compares the distributions of pericenter distances and eccentricity before and after CE evolution. While the final eccentricity distribution is indeed peaked at zero, there is a tail of systems with some residual eccentricity, up to $e_f \sim 0.2$. This is surprisingly similar to the observed distribution, even though our simulations consider CE of a broad range of stellar types, and not just of subdwarf O/B stars or white dwarfs \citep[see Fig. 2 from][]{kruckow2021}.
\REV{Given that the residual eccentricities in our model are very small, the final pericenter distribution does not differ significantly from the one obtained from \textsc{bse}.}

\section{Summary}\label{sec:concl}
In this paper, we presented a new semi-analytic model of CE evolution, which can be incorporated in $N$-body and BPS codes alike. Even though our model is based on simple assumptions, it offers several advantages with respect to previous models. First, it can be easily incorporated into $N$-body codes, enabling the modeling of CE evolution in hierarchical triple and higher multiple stellar systems without the need of running expensive hydrodynamical simulations.
Finally, it can be made consistent with the $\alpha \lambda$ model, widely used in BPS codes, with the advantage of being able to follow the CE inspiral and provide the final eccentricity of the binary.

Even though our model is still mostly phenomenological, i.e., relying on convenient (albeit arbitrary) parametrizations, it lays the foundations for a more predictive semi-analytical model of CE evolution. We explore this direction in our forthcoming work, which will focus on incorporating improved physics into the model, including more realistic density profiles, angular momentum exchange, and mass loss.

\appendix
\section{Mass loss prescriptions}\label{sec:massloss}
For completeness, we write here the equations of motion for adiabatic mass loss or mass gain. If both particles accrete or lose mass isotropically (i.e., with no net linear momentum transfer onto the particles), at rates $\dot{m}_1$, $\dot{m}_2$, we can write the total mass change as $\dot{M} = \dot{m}_1 + \dot{m}_2$. From energy conservation, it is possible to derive the net force per unit-mass acting on the reduced mass system:
\begin{equation}\label{eq:massloss}
	\mathbf{f}= - \frac{\dot{M}}{2M} \mathbf{v}\,.
\end{equation}

We can see that this perturbative force amounts to a linear drag, that is \autoref{eq:forcepowlaw} with $l=1$, $k=0$, and $C=\dot{M}/2M$.
Consequently, the equations of motion follow from Equations~\ref{eq:powerlawkl}--\ref{eq:powerlawkl4}:
\begin{align}\label{eq:adiabmassloss1}
	\dot{a} & = - \frac{\dot{M}}{M} \frac{a}{1-e^2} (1 + e^2 + 2e\cos{\nu})\,,\\\label{eq:adiabmassloss2}
	\dot{e} & = - \frac{\dot{M}}{M} (e + \cos{\nu}) \,,\\\label{eq:adiabmassloss3}
	\dot{\omega} & = - \frac{\dot{M}}{M} \frac{\sin{\nu}}{e}\,.
\end{align}
These equations can be added directly to the drag force equations, once a suitable expression for $\dot{M}$ is provided.

Note that these equations are valid as long as the perturbative force in \autoref{eq:massloss} is relatively small compared to the Newtonian one. If this is not the case, a better description is provided by the impulsive approximation, which assumes instantaneous mass loss \citep{hil80}.

\REV{
Equations~\ref{eq:adiabmassloss1}--\ref{eq:adiabmassloss3} can also be orbit-averaged by integrating them over the mean anomaly (see \autoref{eq:orbitaveraged}, and, e.g., \citealp{hadjidemetriou1963}). The resulting secular equations are:
\begin{align}\label{eq:adiabmasslossave1}
	\langle\dot{a}\rangle & = - \frac{\dot{M}}{M} a \,,\\\label{eq:adiabmasslossave2}
	\langle\dot{e}\rangle & = 0 \,,\\\label{eq:adiabmasslossave3}
	\langle\dot{\omega}\rangle & =0 \,.
\end{align}

In other words, assuming isotropic and adiabatic mass changes, the only net secular effect is the change in the binary semimajor axis, while the eccentricity and the apsidal orientation are not affected. This is in agreement with the bottom panel of \autoref{fig:nbodycomp}, which shows the integrated evolution of the orbital elements for a drag force with $l=1, k=0$. Neglecting the fast oscillations on the orbital timescale, the semimajor axis decreases exponentially (as a straight line in the log-linear plot) while $\omega$ and $e$ remain constant, in agreement with Equations~\ref{eq:adiabmasslossave1}--\ref{eq:adiabmasslossave3}.
}

\section{Orbit-averaged equations}\label{sec:secular}

Equations~\ref{eq:powerlawkl}--\ref{eq:powerlawkl4} express the evolution of the binary orbital elements, including the true anomaly $\nu$, while BPS and triple evolution codes adopt secular equations that average out the dependency on $\nu$. It is not easy to obtain orbit-averaged expressions for generic power-law exponents $l$ and $k$, so we write here the orbit-averaged expressions for the most physically relevant cases.

In the following, the functions ${\rm K}(x)$, ${\rm E}(x)$, and ${\rm \Pi}(x,y)$ are the complete elliptic integrals of the first, second, and third kind, respectively. The orbit-averaged argument of pericenter $\langle \dot{\omega} \rangle$ is zero for all the reported cases.
Finally, to simplify the writing, we make use of the following auxiliary variables:
\begin{align}
	& x = -\frac{4e}{(e-1)^2} \,,\\
	& y = \frac{4e}{(e+1)^2} \,,\\
	& z = \frac{2e}{e+1} \,,\\
	& w = \frac{2e}{e-1}\,.
\end{align}

\begin{figure}[b]
	\centering
	\includegraphics[width=1\linewidth,trim=0.38cm 0.40cm 0.35cm 0.38cm,clip]{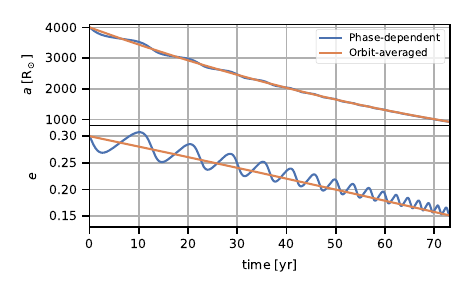}
	\caption{\REV{Comparison between phase-dependent (Equations~\ref{eq:powerlawkl}--\ref{eq:powerlawkl4}) versus orbit-averaged evolution (Equations~\ref{eq:l2k0Analytics}--\ref{eq:l2k0Analytics1}) for the case $l=2$, $k=0$. Top panel: semimajor axis. Bottom panel: eccentricity. Both integrations are carried out with a Dormand-Prince 8th order integrator. In both cases, $\chi_a = 0.05$. }}
	\label{fig:ave_phase}
\end{figure}

\begin{widetext}
\subsection{$f \propto v^2$ ($l = 2$, $k = 0$)}
\begin{align}\label{eq:l2k0Analytics}
    \begin{split}
        \langle \dot{a} \rangle =&\frac{C\sqrt{a\mu}}{(1-e^{2})\pi}\Bigg[(e+1)(e-1)^2 {\rm E}\left(x\right)-(e+1)^2(e-1) {\rm E}\left(y\right)-(7+e^2)(e+1){\rm K}\left(x\right)\\
        &+(7+e^2)(e-1){\rm K}\left(y\right) +4(e+1)^2\Pi\left(w,x\right)+4(e-1)^2\Pi\left(z,y\right)\Bigg]\,,
    \end{split}\\
    \begin{split}\label{eq:l2k0Analytics1}
        \langle \dot{e} \rangle =& \frac{C}{e \pi}\sqrt{\frac{\mu}{a}}\Bigg[(e+1)(e-1)^2 {\rm E}\left(x\right)-(e+1)^2(e-1) {\rm E}\left(y\right)-(3+e^2)(e+1){\rm K}\left(x\right)\\
        &+(3+e^2)(e-1){\rm K}\left(y\right) +2(e+1)^2\Pi\left(w,x\right)+2(e-1)^2\Pi\left(z,y\right)\Bigg]\,.
    \end{split}
\end{align}

\subsection{$f \propto \dfrac{v^2}{r}$ ($l = 2$, $k = 1$)}

\begin{align}
    \begin{split}
        \langle \dot{a}\rangle =& \frac{2C}{(e^2-1)\pi}\sqrt{\frac{\mu}{a}}\Bigg[2(e+1){\rm E}\left(y\right)+2(1-e){\rm E}\left(x\right)+2(e-1){\rm K}\left(y\right)\\
        &-2(e+1){\rm K}\left(x\right)+(e+1)^2\Pi\left(w, x\right)+(e-1)^2\Pi\left(z, y\right)\Bigg]\,,
    \end{split}\\
    \begin{split}
        \langle \dot{e}\rangle =& -\frac{2C}{a e \pi}\sqrt{\frac{\mu}{a}}\Bigg[(e+1){\rm E}\left(y\right)+(1-e){\rm E}\left(x\right)+2(e-1){\rm K}\left(y\right)\\
        &-2(e+1){\rm K}\left(x\right)+(e+1)^2\Pi\left(w, x\right)+(e-1)^2\Pi\left(z, y\right)\Bigg]\,.
    \end{split}
\end{align}

\subsection{$f \propto \dfrac{v^2}{r^2}$ ($l = 2$, $k = 2$)}
\begin{align}
    \begin{split}
        \langle \dot{a}\rangle =& -\frac{2C}{3a(e^2-1)^2\pi}\sqrt{\frac{\mu}{a}}\Bigg[-(e^2+1)(e-1){\rm E}\left(x\right)+7(e^2+1)(e+1){\rm E}\left(y\right)\\
        &+(e^2-1)(e+1){\rm K}\left(x\right)-(e^2-1)(e+1){\rm K}\left(y\right)\Bigg]\,,
    \end{split}\\
    \begin{split}
        \langle \dot{e}\rangle =& \frac{C}{3a^2 e (e^2-1)^2\pi}\sqrt{\frac{\mu}{a}}\Bigg[-(e^2+1)(e-1){\rm E}\left(x\right)+(13e^2+1)(e+1){\rm E}\left(y\right)\\
        &+(e^2-1)(e+1){\rm K}\left(x\right)-(e^2-1)(e+1){\rm K}\left(y\right)\Bigg]\,.
    \end{split}
\end{align}
\end{widetext}

\REV{In \autoref{fig:ave_phase} we compare the numerical integration with the phase-dependent equations (Equations~\ref{eq:powerlawkl}--\ref{eq:powerlawkl4} with $l=2$, $k=0$) and the orbit-averaged counterparts (Equations~\ref{eq:l2k0Analytics}--\ref{eq:l2k0Analytics1}). As expected, the orbit-averaged evolution does not exhibit any oscillations on the orbital timescale, but represents the average value of semimajor axis and eccentricity over one orbit.
}

\REV{
\section{Evolution of the eccentricity with dynamical friction forces ($l=-2$)}\label{sec:lnegative2}
In this section we analyze the evolution of the eccentricity for $l=-2$. This case corresponds to the dynamical friction force in collisionless and collisional media, wherein a massive body moving in sea of smaller bodies is slowed down by the overdensity that forms in its wake.
We note that, during the finalization of this manuscript, \citet{slogien2022} posted an analysis of this case using a simplified toy model, finding that $k=3$ is the boundary between the increase and decrease in eccentricity. Our results agree with their estimate, and in this Appendix we provide a complete explanation grounded in perturbation theory.

\begin{figure}[b]
	\centering
	\includegraphics[width=1\linewidth,trim=0.15cm 0.0cm 0.42cm 0.38cm,clip]{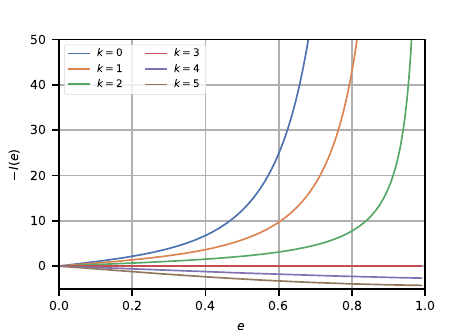}
	\caption{Values of the integral $-I(e)$ that arises when orbit averaging \autoref{eq:edotnegl2}, that is the rate of change in eccentricity for a dynamical friction force ($l=-2$). Each curve corresponds to a different value of $k$. For $k=3$, $\langle \dot{e} \rangle = 0$ and the derivative of the eccentricity changes sign, from positive ($k>3$) to negative ($k<3$). }
	\label{fig:elliptint}
\end{figure}

We begin our analysis by considering that for $l=2$, \autoref{eq:powerlawkl2} reads as:
\begin{equation}\label{eq:edotnegl2}
	\begin{split}
	\dot{e} = & -2C\mu^{-\frac{3}{2}} a^\frac{3-2k}{2} (1-e^2)^\frac{3-2k}{2}\\& (1+e\cos{\nu})^k (1+e^2+2e\cos{\nu})^{-\frac{3}{2}}(e+\cos{\nu})\,.
	\end{split}
\end{equation}
After applying the orbit-averaging technique from \autoref{eq:orbitaveraged}, the integrand function obtains the following form:
\begin{equation}
	\begin{split}	
	\frac{\dot{e}}{2\pi} \frac{dM}{d\nu} = & h(C,\mu,a) (1-e^2)^{3-k} (1+e\cos{\nu})^{k-2}\\& (1+e^2+2e\cos{\nu})^{-\frac{3}{2}}(e+\cos{\nu})\,,
	\end{split}
\end{equation}
where
\begin{equation}
	h(C,\mu,a)=-\frac{C\mu^{-\frac{3}{2}} a^\frac{3-2k}{2}}{\pi}\,.
\end{equation}
After carrying out all the constant terms, the integral in $\nu$ contains only
\begin{equation}
	\begin{split}	
			I(\nu,e) = & (1+e\cos{\nu})^{k-2} (1+e^2+2e\cos{\nu})^{-\frac{3}{2}} (e+\cos{\nu})
	\end{split}
\end{equation}

Unfortunately, we were not able to find a generic solution to the integral
\begin{equation}\label{eq:elliptintegral}
	I(e) = \int_{0}^{2\pi} I(\nu,e) d\nu
\end{equation}
for arbitrary values of $k$. It is however possible to derive analytic solutions for fixed values of $k$. We derived the analytic expressions of $\langle \dot{e} \rangle$ for $k=1,2,3,5$, and show the integral $-I(e)$ for different values of $k$ in \autoref{fig:elliptint}. Here we quote only the complete expression for $k=3$:
\begin{equation}
	\begin{split}
		\langle \dot{e} \rangle = & \frac{2 C}{\pi  a^{3/2} (e+1)^2 \mu ^{3/2}} \Bigg[(e-1) {\rm E}\left(x\right)+(e+1) {\rm E}\left(y\right)\Bigg]\,,
	\end{split}
\end{equation}
where we have used the notation of the previous section. It can be shown that for $\left|e\right|<1$, $(e-1) {\rm E}\left(x\right)= - (e+1) {\rm E}\left(y\right)$ and consequently $\langle \dot{e} \rangle = 0$ for $k=3$.

}

\begin{acknowledgments}
We would like to thank the anonymous referee for multiple useful comments and suggestions that helped to improve this work. \REV{A.A.T. wishes to thank Evgeni Grishin for helfpul comments and criticism.}
This research was supported by JSPS Grants-in-Aid for Scientific Research (17H06360, 19K03907, 21K13914).
SR acknowledges funding from STFC Consolidated Grant ST/R000395/1 and the European Research Council Horizon 2020 research and innovation programme (Grant No. 833925, project STAREX).
GI acknowledges financial support from the European Research 
 Council for the ERC Consolidator grant DEMOBLACK, under contract no. 
 770017. 
This work was supported by NOVA.
\end{acknowledgments}

\bibliography{totalms.bib,additionalref.bib}

\end{document}